\documentstyle[aps,pre,eqsecnum,epsf,psfig,twocolumn]{revtex}
\input epsf

\def\sep{\ell}
\def\delroro{\frac{\Delta \rho}{\rho_0}}
\def\subequations{\begin{mathletters}}
\def\endsubequations{\end{mathletters}}

\def\dddk{ \frac{\dt k^2}{\sok}}
\def\tpd{{(2 \pi)}^d}

\def\half{\frac{1}{2}}

\def\be{\begin{equation}}
\def\ee{\end{equation}}
\def\ber{\begin{eqnarray}}
\def\eer{\end{eqnarray}}
\def\lav{\left <}
\def\rav{\right >}
\def\ddt{\frac{\partial}{\partial t}}
\def\drdt{\frac{\partial \rho}{\partial t}}

\def\ddz{\frac{\partial }{\partial z}}
\def\dfdr{\frac{\delta F}{\delta \rho}}

\def\bx{{\bf x}}
\def\br{{\bf r}}
\def\bxp{{\bf x'}}
\def\bj{{\bf J}}
\def\bz{{{\bf z}}}
\def\muc{\psi}
\def\muo{{\mu}_0}
\def\mub{\left({\frac{\mu}{\mu}_0}\right)}
\def\mutwo{{\mu}_2}
\def\r0{{\rho}_0}
\def\dr{\delta \rho}
\def\drsq{{\dr}^2}
\def\ft{\tilde F}

\def\dt{\tilde D}
\def\aa{  {\left(  \frac{\partial \mu}{\partial \rho}  \right)}_{\r0}}
\def\bb{{ \left(  \frac{{\partial}^2 \mu}{\partial {\rho}^2}  \right) }_{\r0}}
\def\lr{\left( ln \left( \frac{\rho}{\rho_0} \right) \right)}
\def\bg{{\bf G}}
\def\bgone{{\bf G_1}}
\def\bgtwo{{\bf G_2}}
\def\bgthree{{\bf G_3}}
\def\bgfour{{\bf G_4}}
\def\bk{{\bf k}}
\def\bmk{{-\bf k}}
\def\bkp{{\bf k'}}
\def\bkmp{{\bf k - \bf p}}

\def\bkone{{\bf k_1}}
\def\bktwo{{\bf k_2}}
\def\bkthree{{\bf k_3}}
\def\bkfour{{\bf k_4}}
\def\bp{{\bf p}}
\def\si{\sum_{i=1}^3}

\def\ix{\int d \bx}

\def\ik{\int \frac{d {\bk}}{{(2 \pi)}^d} }
\def\ip{\int \frac{d {\bp}}{{(2 \pi)}^d} }
\def\ipp{\int \frac{d {\bp}'}{{(2 \pi)}^d} }
\def\phik{\phi (\bk)}
\def\phikone{\phi({\bk}_1)}
\def\phiktwo{\phi ({\bk}_2)}
\def\phikthree{\phi ({\bk}_3)}
\def\phimk{\phi (- \bk)}
\def\phip{\phi (\bp)}
\def\phikmp{\phi (\bk - \bp)}
\def\sok{S_0(k)}
\def\sop{S_0(p)}

\def\sopmk{S_0(|\bp - \bk|)}
\def\sokmp{S_0(|\bk - \bp|)}
\def\kisq{{k_i}^2}
\def\kiz{{k_i}_z}
\def\kisqbysk{\frac{\kisq}{\sok}}
\def\kizbysk{\frac{\kiz}{\sok}}
\def\lll{ i \lambda k_z}
\def\vst{{v_{\mbox{\footnotesize Stokes}}}}
\def\bg{{\bf G}}

\def\kineticcoef{\Gamma_0}
\def\ikone{\int \frac{d {\bkone'}}{{(2 \pi)}^d} }
\def\iktwo{\int \frac{d {\bktwo'}}{{(2 \pi)}^d} }
\def\ikthree{\int \frac{d {\bkthree'}}{{(2 \pi)}^d} }
\def\half{\frac{1}{2}}
\def\be{\begin{equation}}
\def\ee{\end{equation}}

\def\t1{{\nabla}_{\bot}  u_z}

\begin{document}
\draft
\title{Correlations and Freezing in Steadily Settling Suspensions}
\author{Rangan Lahiri\cite{a}} 
\address{Theoretical Physics Group, Tata Institute of Fundamental
Research, Homi Bhabha Road, Mumbai 400 005, India.}
\author{Sriram Ramaswamy\cite{b}}
\address{Department of Physics, Indian Institute of Science,
Bangalore 560 012, India.}
\date{version of October 8 1998 draft 5; printed \today}
\maketitle

\begin{abstract}
We study the liquid-solid transition in a  
collection of interacting particles moving through  
a dissipative medium under the action of a constant, 
spatially uniform external force,
e.g. a charge-stabilized suspension in a fluidized bed or  
a flux-point lattice moving through a thin, current-carrying 
slab of type II superconductor. 
The mobility of a given region in these systems is in 
general a function of the local concentration.
We show that the structure factor peak is suppressed in 
an anisotropic manner as a
result of this effect, resulting in a shift of the
crystal-liquid phase boundary towards the crystal side.
A nonequilibrium phase diagram is presented.
\end{abstract}


%%%%%%%%%%%%%%%%%%%%%% from introsp.tex %%%%%%%%%%%%%%%%%%%%%%%%%%%%%%%%%%

\section{introduction and results}
\label{introinpaper}
How a liquid freezes to form a solid is a problem of
longstanding interest to physicists.
Freezing at thermal equilibrium can now be said to be 
quite well understood in terms of a first-principles 
order-parameter theory \cite{RY}.  
The effect of steady nonequilibrium driving on the 
freezing transition has been studied in the context of 
{\em sheared} colloids \cite{shearmelting,Stevens,rrbagchi,rlshear}.   
A uniformly driven state, with no gradients on the average, 
as would arise in steady sedimentation \cite{russell,introsed} 
or the fluidized bed 
\cite{rutgersthesisintro,introhappel,rutgerspreintro}
is conceptually simpler: unlike in 
shear flow, an order-parameter description of the mean state 
of the driven crystal is straightforward. Nonetheless,  
we know of no studies, experimental or theoretical, 
of the effect of sedimentation on the liquid-solid phase boundary 
in a suspension settling uniformly under gravity, or of the 
freezing of moving flux lattices in the {\em absence} of quenched disorder. 

A study of freezing in the presence of uniform driving is also of 
interest as a problem in the general class of driven diffusive systems.
Most of the investigations in this field \cite{ddlg}
are either away from any equilibrium phase transition or 
in the neighbourhood of a second order phase transition.
The effect of uniform driving on a first order phase transition has
received very little attention \cite{muktish}; the freezing transition, 
in particular holds interesting possibilities since the ordering is 
at non-zero wavenumber.

In this paper we consider a collection of interacting particles 
in a liquid state, i.e., with strong short-range translational 
correlations, and
ask what happens to the liquid-solid transition in the presence
of an imposed steady mean translational motion
of the particles with respect to the ambient medium.
Our model should apply not only to charge-stabilized colloidal 
suspensions \cite{collbasic},
but also to the effect of a steady current on the freezing
transition of flux-point liquids \cite{fluxbasic}
in thin slabs of clean type II superconductors.
The main ingredient in our theory is the concentration
dependence of the mobility of particles in a suspension,
or of the flux-points in a superconductor.
In colloidal suspensions, 
this dependence arises as a result of the
hydrodynamic interactions between particles,
while  in the flux-liquid, it is the electromagnetic
interaction that is primarily responsible for the effect.
Although our approach is fairly general, 
we use parameter values appropriate to the specific case 
of sedimenting colloidal suspensions in the rest of the paper.

Our main results are as follows: As the mean concentration of the 
suspension is increased,
the hydrodynamic interaction causes it to settle more slowly
\cite{rutgersthesisintro,xue,hamhom}.
We show that this leads to a nonlinearity in the equations 
of motion whose effect is to reduce the height of the structure 
factor relative to its equilibrium value. The suppression is 
strongly anisotropic, largest for wavevectors along the direction 
of mean drift, and vanishing for wavevectors normal to it. The reduction 
is most pronounced at the location of the peak, which means the 
strong short-range correlations in the liquid, which are the 
precursor of translational order, are suppressed. This inevitably  
leads to a shift of the phase boundary towards the crystal side 
(i.e., favours the liquid over the crystalline phase),  
by an amount proportional to the square of the P\'eclet number, 
in a manner reminiscent of a proposed mechanism \cite{rrbagchi} 
for shear-induced melting.
This nonequilibrium phase transition now assumes
additional importance in the wake of recent experiments on 
charged fluidized beds
\cite{rutgersthesisintro,introhappel,rutgerspreintro,xue}
where it should be possible to test our predictions. Constraints 
on the types of systems to which our theory should apply are 
discussed at the end of section \ref{sec:conclude}.

The paper is organised as follows.
Section \ref{introfreezing} gives an overview of equilibrium freezing.
The equations describing the motion of a sedimenting suspension
are derived in section \ref{freezingeomsection}.
In section \ref{freezingphasetrans}
we discuss in a general setting the prescription used in this
paper to describe a nonequilibrium phase transition.
The perturbative calculation leading to the equal time correlation
function in the driven system is the content of section 
\ref{freezingperturbation},
and our numerical results for the structure factor are given in
section \ref{freezingstrucfac}.
In section \ref{freezingfreezing} we derive our results for the freezing transition,
and conclude in section \ref{sec:conclude}.

\section{Equilibrium freezing}
\label{introfreezing}

The equilibrium phase-transition from a liquid to a crystal is the spontaneous
appearance of a static, spatially periodic modulation in the density
field of the liquid.
An order parameter theory of this transition \cite{alexander,chailub}.
is obtained by writing
down the free-energy cost of such a modulation as a functional of the
density field.

The free energy cost of an inhomogeneous density field $\rho(\bx)$
relative to the uniform liquid state of density $\rho_0$ is given by
\cite{RY}
\ber
\ft 
&=& 
F/k_B T 
\nonumber\\
&=&
% sriram -rrbagchi,rrbagchi,rrbagchi,
\half \ix\/ c(|\bx - \bxp|) (\rho(\bx) - \rho_0) (\rho(\bxp) - \rho_0)
\nonumber\\
&& + \ix  \left[ 
\rho(\bx) \mbox{ln} \frac{\rho(\bx)}{\rho(0)} -(\rho(\bx) - \rho_0)
\right],
\label{ryf}
\eer
where $c(r)$ is the direct correlation function of the liquid.
The second term in (\ref{ryf}) is an ideal  gas entropy, 
whereas the  first comes from correlations in the interacting system.
Expanding the logarithm and stopping at the second order 
in the density fluctuation
$\phi = \rho - \rho_0$,
gives the quadratic part of the free-energy 
in terms of the Fourier modes $\phik$ as:
\be
\ft =\half \ik \frac{1}{\rho_0 \sok} {|\phi(\bk)|}^2
\ee
in terms of the dimensionless structure factor
\be
\sok = \frac{1}{1 - \rho_0 c(k)}.
\ee
The density correlation function at equilibrium is related to the 
structure factor by the relation
\be
\lav \phik \phi(- \bk)  \rav
=
V \rho_0 \sok
\ee
where V is the system volume.

All information about temperature and mean density is encoded in the
liquid-state direct correlation function $c(r)$.
For a given $c(r)$, $\rho(r)$ corresponding to the thermodynamically
stable state, within mean-field theory, is the global minimum of $F[\rho]$,
obtained in principle by solving $\frac {\delta F}{\delta \rho} = 0$,
and choosing that solution for which $F$ is the smallest.
At low densities and high temperatures, the global minimum
is $\rho(r) = \rho_0$.
For sufficiently dense, cold systems, this uniform liquid state is
unstable relative to a spatially periodic $\rho (r)$, the crystal.
Terms in $F$ of order higher than quadratic in $\phi(r) = \rho(r) - \rho_0$
are clearly essential to give such a solution.
Although the best theories of freezing \cite{RY}
go the whole log, we shall sacrifice accuracy in favour of simplicity,
and use a quartic polynomial approximation to $F$. 
Since we are looking for periodic density waves, it is convenient to 
work with $\{ \phi_{\bg} \}$, the weights of the Bragg peaks at the reciprocal
lattice vectors $\bg$, defined by
\be
\phik = \sum_G \phi_\bg \delta(\bk - \bg).
\ee

Since the peak of the  liquid-state structure factor at its
first maximum $ G = 2 \pi/a$ is much higher than the subsequent peaks,
we restrict ourselves to Fourier amplitudes of $\phi$
close to this peak.  
This is equivalent to considering only the first shell of reciprocal
lattice vectors for the crystal.
Only the form of the S(k) near the peak matters then, and
$S^{-1}(G)$ can be replaced by a Lorentzian
\be
S^{-1}(G)  = a + \sigma (G^2 - {k_0}^2 )^2 
\ee
so that the peak height is
\be
S(k_0)  = a^{-1}.
\ee
The Landau Free energy per particle is then given by
\ber
\frac{\ft}{N}  &=& 
\frac{1}{2} \sum_\bg 
\left[a + \sigma (G^2 - {k_0}^2 )^2 \right]
\frac{\displaystyle \phi_\bg \phi_{-\bg}}{\displaystyle {\rho_0}^2}
\nonumber
\\
&&- \frac{1}{3!} B {\rho_0}^2 
\sum_{\bgone,\bgtwo,\bgthree}
\frac{\displaystyle \phi_{{\bf G}_1} \phi_{{\bf G}_2} \phi_{{\bf G}_3} }
{\displaystyle {\rho_0}^3}
\,
\delta_{{\bf G}_1 + {\bf G}_2 + {\bf G}_3,0} 
\nonumber \\
&&+ \frac{1}{4!} C {\rho_0}^3 
\sum_{\bgone  \ldots \bgfour}
\frac{\displaystyle \phi_{{\bf G}_1} \phi_{{\bf G}_2} \phi_{{\bf G}_3} \phi_{{\bf G}_4} }
{\displaystyle {\rho_0}^3}
\delta_{{\bf G}_1 + {\bf G}_2 + {\bf G}_3 + {\bf G}_4 ,0}. 
\nonumber\\
\label{flandauintro2}
\eer

If $\sigma {G_0}^4 \gg a$, then
free energy costs strongly discourage the Fourier amplitudes to stray
from the peak of the structure factor, i.e $|{\bf {G}}| = k_0$.
The free energy can then be expressed as a function of a single scalar
order parameter
\begin{equation}
r = n^{1/2} \left(\rho_\bg/\rho_0\right)
\label{opr}
\end{equation}
where $n\/$ is the number of reciprocal lattice vectors in
the set (i.e. 6 for a triangular lattice, 12 for bcc etc).
Note that at nonzero wavevector $\bg$, 
$\phi_\bg$ and $\rho_\bg$ are identical.
The free energy per particle per $k_B T$ is thus a polynomial
\be
% sriram 
f  = \frac{F}{N k_B T} = \frac{1}{2} a r^2 - \frac{1}{3!} b r^3 + \frac{1}{4!} c r^4,
\label{flandauintrooneop}
\ee
with
\subequations
\begin{eqnarray}
b  &=& 2 p n^{-1/2} {\rho_0}^2 B
\label{landauparameterb}
\\
c  &=& 6 \left[ 1 + \frac{q}{n} \right] {\rho_0}^3 C.
\label{landauparameterc}
\end{eqnarray}
\label{landauparameters}
\endsubequations
Here $p$ and $q$ are numbers that depend on 
the type of lattice being considered:
$p$ is the number of such triplets or triangles
that each RLV belongs to, and $q$ is the number of non-planar diamonds
that the RLVs can form among themselves \cite{chailub}.
The value of $a$ at which the transition occurs is found by demanding
that the free energy be zero at the crystalline minimum.
This leads to the value of $a$ at the transition,
\be
a_c = \frac{b^2}{3 c}.
\ee

The cubic term favours lattice structures in which
the RLVs form triangles in sets of three.
The candidate lattices, within the  one order parameter theory, are
the ones whose RLVs correspond to edges of regular polyhedra or 
polygons with triangular faces, i.e. planar triangular (whose 
RLVs form a pair of triangles),
face centred cubic (octahedron) and icosahedral (icosahedron) lattices.
The theory predicts that  all crystals should be triangular
lattices in two dimensions and body centred cubic 
(which has a face centred cubic reciprocal lattice) in three dimensions.
Given the lack of  sophistication, the theory is reasonably successful
as almost all two dimensional crystals are triangular lattices, and
many, if not all elemental solids in three dimensions are indeed BCC, at
least close to freezing \cite{alexander}.

The  theory discussed above has several drawbacks: it neglects all
structure in S(k) except  that  at the  peak and assumes that all
wavevectors  have the same order parameter amplitude.
These assumptions can be relaxed and a more complete theory 
written down, but the Landau expansion
itself is questionable, in fact  inaccurate, since the  order parameter
jump at freezing is large.
It would doubtless be better to generalize the first-principles
density-wave approach of Ramakrishnan and Yussouff 
\cite{RY}
to include the effects of sedimentation.
In the present work, however, we have 
attempted only a nonequilibrium generalizations of
the simple and intuitive Landau-Alexander-McTague theory,
as a first step in taking into account the effect of sedimentation
on freezing.
An extension of our approach to more realistic freezing theories
should be fairly straightforward.
 
\section{equations of motion}
\label{freezingeomsection}
The frictional force on a spherical particle of
radius $R$ moving slowly in a fluid with shear viscosity $\eta$ 
with a uniform speed $v$ is given by \cite{stokes2}
\be
{\cal F} = 6 \pi \eta R v
\label{stokeslaw}
\ee
The dimensionless parameter which gives a measure of the
relative importance of sedimentation with respect to Brownian motion is
the P\'eclet number, defined as the ratio of the 
time taken by the particle to diffuse a characteristic distance $\sep$
to the time taken to settle the same distance under gravity, without diffusing.
In terms of the mean settling speed $\vst$ of the particle, the P\'eclet number
is given by
\be
Pe =\frac{v_{\mbox{\footnotesize Stokes}} \sep}{D}.
\ee
The appropriate dimensionless measure (which we will continue 
to call the P\'eclet number)  of the relative 
importance of drift versus diffusion for this problem, 
will turn out to be proportional to the P\'eclet number defined 
at the scale of an interparticle spacing, but will 
incorporate certain collective effects whose nature 
will become clear below after we construct our equations of motion.

Although the single particle problem was solved in 1851 \cite{stokes2},
the sedimentation speed of a suspension was theoretically obtained
as recently as 1972 \cite{introbatchelor1},
and that too in the dilute limit.
The speed of a strictly random suspension of volume fraction $\phi$ 
turns out to be {\em less} than that of a single sphere:
\be
v_{\phi} = (1 - 6.55 \phi + {\cal{O}}(\phi^2)
)  \/ v_{\mbox{\footnotesize Stokes}}
\label{hinderfluid}
\ee
Thus, at a volume fraction of $10 \%$, the settling speed is 
reduced to $35 \%$ of the Stokes value.
It has been also observed experimentally \cite{xue,hamhom}
that the mean settling speed of a suspension is less than
that of a single particle, and a decreasing function of the
concentration of suspended particles.
In other words, the mobility of a region is a function
of the local concentration. 

The sedimenting suspension that we speak of is an idealized
bottomless tube of infinite height containing  a solvent
in which there is a steady downward
flow of particles (denser than the solvent).
An `infinite' tube is realized in real life by means of
a setup known as a fluidized bed 
\cite{rutgersthesisintro,introhappel,rutgerspreintro,xue},
a vertical tube through which fluid is pushed up at a steady rate
by an externally maintained pressure head, with particles suspended 
stably on the average by the balance between gravity and the viscous drag 
of the upflow of fluid. 
The problem of the steadily sedimenting particle in the comoving
frame of reference is identical, in the infinite container 
limit, or far from the walls of the container, to the fluidized bed problem 
in the lab frame.
For very dilute solutions,
the  reduced weight of the  particle (i.e. the  weight minus the
buoyant force) is balanced by the single particle Stokes friction.
At higher concentrations, hydrodynamic interactions between the  spheres
\cite{vonintro,laddintro}
becomes important, leading to 
a hindered settling speed, the effect of which we investigate.

Consider now a collection of interacting particles,
on the average in a liquid state with uniform number density $\rho_0$,
moving through a frictional medium. 
In the absence of external forces (such as gravity)
the number density $\rho(\bx)$ follows the equation:
\be
\drdt = \kineticcoef {\nabla}^2 \dfdr + \eta,
\label{eomeq}
\ee
where $\kineticcoef$ is the collective diffusion constant, $\eta$ is a 
Gaussian white noise consistent with thermal equilibrium:
\be
\left< \eta ( \bx , t) \eta ( \bxp , t') \right>
=
2 k_B T \kineticcoef {\nabla}^2 \delta (\bx - \bxp) \delta (t - t'),
\label{noiseeq}
\ee
and $F$ is the free-energy functional defined in \ref{ryf}.
This equation is simply the equation of continuity 
\be
\drdt = - \nabla . \bj
\ee
with a current with two pieces: a mean
$ \left< \bj \right> = - \kineticcoef \nabla \muc$
and a fluctuating piece ${\bf f}$, uncorrelated in space and time.
The chemical potential $\muc$ is related to the Gibbs free energy $F$
by the thermodynamic relation
$
\muc = \dfdr
$
which gives equation (\ref{eomeq}) with a conserving noise
$\eta = - \nabla. {\bf f}$.
Note that the equation of motion \ref{eomeq} combined with the 
noise statistics (\ref{noiseeq}) will take the system to an equilibrium
state with a probability distribution
\be
P \sim e^{- F / k_B T}.
\ee

In the presence of a constant external force, which we take for specificity
to be a uniform gravitational field 
$
{\bf g} = g  {\hat{\bf z}},
$
the particles sediment with a steady mean drift velocity
$
{\bf v} = {\mu}( \rho ) m {{\bf g}}\/
$
\cite{kynch}, where $\mu$ is the concentration-dependent mobility, 
which in the low density limit, is expected to go to the Stokes value 
$
\frac{1}{6 \pi \eta R}\/.
$
The current then acquires an extra piece $ \rho {\bf v}$ leading to 
an additional term of the form
\be
 - m g \ddz [ \rho \mu ( \rho ) ]
\label{extraterm}
\ee
on the right hand side of equation (\ref{eomeq}).
Here $mg$ is the reduced or Archimedean weight of the solute particles,
\be
mg = m_0g - m_{\mbox{\footnotesize{displaced}}}g,
\ee
with $m_0$ being the mass of the particle, and 
$m_{\mbox{\footnotesize {displaced} }}$ the mass of solvent displaced 
by submerging it.

Let us now consider concentration fluctuations in the suspension around the
mean density ${\rho}_0$:
$
\rho = \r0  + \dr.
$
Taylor expanding the mobility about the mean value $\muo$
to second order, we find the equation of motion for the 
density field to be 
\ber
\frac{\partial }{\partial t} \dr &+& mg ( \muo + \r0 \aa ) \ddz \dr 
\nonumber\\
&=&
\kineticcoef {\nabla}^2 \dfdr - \mutwo \ddz  \drsq + \eta,
\label{eom0}
\eer
where 
\begin{eqnarray}
\mutwo & = & m g\; \left[ \aa + \r0 \bb \right]
\nonumber
\\
& =& \frac{\muo m g}{\r0} \;  
{\left[ \frac{{\partial}^2 { \mub}}{\partial {\lr}^2} \right]}_{\r0}.
\label{mutwodef}
\end{eqnarray}
The left hand side of (\ref{eom0})
is simply the Eulerian derivative for a reference frame
moving with a velocity
\be
{\bf v}_{\mbox {\footnotesize{frame}}} = g ( \muo + 
({\frac{{\partial} {(ln \mu)}}{\partial {\rho}})}_{\r0})
\ddz \left( \dr \right) \; {\hat \bz}.
\ee
In the rest of this paper, we will do all our calculations in this
reference frame, so that the left hand side of (\ref{eom0}) can be
replaced by a time derivative.
Note that this frame is {\em not} the frame in which the mean velocity of the
particles is zero.
For a colloidal suspension,
since $\mu$ decreases on increasing concentration, the
frame moves with a velocity less than the mean velocity.
Thus one would see a net drift along gravity in this frame.
However, the quantities of interest to us, i.e., equal time correlation
functions, are invariant under a Galilean transformation, so it
doesn't matter in which frame we do our calculation
\cite{fnfreezingframes}.

Before beginning the calculations, we change variables to the
non-dimensionalized fluctuation of the concentration from the mean,
$\phi = \frac{\rho - \r0}{\r0}$,
and the dimensionless free energy 
$\ft = F / k_B T$,
in terms of which the equation of motion in the comoving frame becomes
\subequations
\begin{eqnarray}
\frac{\partial \phi}{\partial t} &
=&
\dt {\nabla}^2 \frac{\delta \ft}{\delta \phi} 
+ \lambda \ddz  {\phi}^2 + \eta,
\label{eomphi}
\\
\left< \eta ( \bx , t) \eta ( \bxp , t) \right>&
=&
-\dt \nabla^2 \delta (\bx - \bxp) \delta (t - t'),
\label{eomnoise}
\end{eqnarray}
\label{freom}
\endsubequations
where
\be
\dt =\frac{k_B T \kineticcoef}{{\r0}} 
\ee
is the collective diffusion constant, and 
\be
\lambda = - \muo m g 
{\left[ \frac{{\partial}^2 { \mub}}{\partial {\lr}^2} \right]}_{\r0}
\ee
is a velocity scale set by the Stokes velocity modified by the 
concentration-dependence of the
mobility. Note that the piece  $\muo m g $ outside the brackets is
the mean settling speed.  
The term within the brackets is a dimensionless number that 
vanishes in the dilute limit.
For sedimenting suspensions, the mobility is a decreasing function of
concentration, so $\lambda > 0.$
Since $\dt$ and $\lambda$ are a diffusion constant
and a velocity respectively, together they give the 
appropriate dimensionless coupling (which we will call 
the P\'eclet number $Pe$) for the problem:
\subequations
\begin{eqnarray}
Pe & = & \frac{\lambda \sep}{ \dt}
\label{peclet1}\\
& = &\left( \frac{\muo m g \sep {\rho_0}^2}{k_B T \kineticcoef} \right)
{\left[ \frac{{\partial}^2 { \mub}}{\partial {\lr}^2} \right]}_{\r0},
\label{peclet2}
\end{eqnarray}
\label{peclet}
\endsubequations
where $\sep$ is the average interparticle spacing in the suspension.
Note that we are using $\sep$ rather than the particle size as the
length scale to define the P\'eclet number.
This is reasonable since we are considering collective
rather than single particle effects.
In terms of the single particle diffusion constant $D_0$, which
is related to the bare mobility by the Einstein relation
$D_0 = k_B T \mu_0$, 
the P\'{e}clet number can be written as
\be
Pe  = 
\left(\frac{D_0}{\dt} \right)
\left(\frac{\sep}{R} \right)
{\left[ \frac{{\partial}^2 { \mub}}{\partial {\lr}^2} \right]}_{\r0}
\;
\left(\frac{mg R}{k_B T} \right).
\label{peclet3}
\ee
The last piece (in parentheses) in (\ref{peclet3})
is the particle-size P\'eclet number for a single sphere, which is simply
the ratio of the
gravitational potential energy drop across the size of the particle
to the thermal energy $k_B T$.
The other three terms are dimensionless numbers that bring in
the collective effects.

The hydrodynamic interactions appear through the third piece in 
\ref{peclet3} (in square brackets).
It should be noted that this term is indeed nonzero even if the mobility is a 
linear function of concentration.

We pause here to do some power-counting.
The change in mobility due to a number density fluctuation depends
on the change in {\em volume fraction}. 
The Taylor expansion of mobility, for example, is given by 
$\mu = \mu_0 \left[ 1  -  \alpha R^d ( \rho - \rho_0) + \ldots \right]$
in $d$ dimensions, 
where $\alpha$ is a dimensionless number that can be found from experiments.
To the lowest order in the polyball size R, therefore, the piece in
square brackets in \ref{peclet3} is proportional to $R^d$.
The P\'eclet number thus goes as $m \/ R^d$ 
for particle of mass $m$ and radius $R$.
As $R \longrightarrow 0$, the hydrodynamic interaction between the spheres
goes away, and the P\'eclet number goes to zero as $R^{d}$ if we keep
the mass fixed, and as $ R^{2 d}$ if we change the radius of the
particle keeping the density fixed.

The equation of motion (\ref{freom}) of the density fluctuation is the
equilibrium dynamical equation with an additional nonlinearity of the 
Burgers type \cite{kpz}.
Qualitatively, the nonlinear term may be understood in terms of
a lattice gas model \cite{ddlg}, where particles are
driven in one direction by a constant field, but a particle
is allowed to hop to the next site only if it is empty
(no double occupancy).
This microscopically models a mobility depending on the local
density --  in regions where the density is large, the probability
of hopping being obstructed by the presence of other particles is
larger, and hence the mobility smaller.
The current at site $n$ is then proportional to $\rho_n ( 1 - \rho_{n+1})$,
where $\rho_n$ is the discretized density in a bin at $n$.
On coarse graining this will give two terms:
a term $\sim \phi$, which can be removed by a Galilean transformation;
and another $\sim - \phi^2$, which persists even in the comoving frame.
%The lack of Galilean invariance of the equations of motion
%is due to the presence of a fixed lattice, with respect to which one
%frame is static whereas the other is moving.

Before moving on to use (\ref{freom}) to calculate the correlations in
the suspension, let us briefly discuss the limitations of our model.
First, we have only considered one effect of sedimentation, namely
the dependence of concentration on mobility.
This is a coarse-grained description of the hydrodynamic interactions,
reliable only at long wavelengths.
A better theory should consider, for example, gradient-dependence in
the mobility, i.e., the diffusivity at wavenumber $k$ should have corrections
of ${\cal O} (k R)$ for a suspension of particles of size $R$.
Since we are interested in ordering phenomena at finite wavevectors $\sim k_0$, 
our theory should give reliable results for $k_0R << 1$
which is satisfied, for example, for strongly charge stabilized colloids 
where the ordering length scale ${k_0}^{-1}$ 
is of order the screening length, $\xi \gg R$. In addition, we have 
ignored the dependence of the mobility on more detailed aspects 
of the local distribution of particles, e.g., its anisotropy, 
a point which we comment on at the end of this paper.

Secondly, we have also ignored the concentration-dependence of the diffusion
constant $\dt$ in (\ref{eomphi}). 
Strictly speaking, $\dt$ should have the same concentration dependence
as $\mu_0$, since the two are related by the Einstein formula.
A correct formulation  of the problem with $\dt(\phi)$
would require a multiplicative
noise (of the form $\nabla.[\sqrt{\tilde{D}(\phi)} \; {\bf f}]$ 
where ${\bf f}$ is a spatiotemporally white noise with unit variance)
to make the equations of motion consistent with equilibrium.
This problem is technically much more difficult
\cite{kumaran,srirammult}, 
and we believe it shouldn't change the results dramatically for the following reason.
A concentration-dependent diffusion constant coupled with a consistent
noise of the above type is an {\em equilibrium} effect, and an (effective) concentration
independent $\dt$ with ordinary Gaussian noise leads to the same
equilibrium distribution.
The nonlinear term in (\ref{eomphi}) is the {\em only driving term}
arising from concentration dependence.
In any case,
(\ref{freom})
is the simplest driven-diffusive model with a liquid-solid transition
and the correct equilibrium limit, and thus merits study in its own right.

Lastly, we have also ignored the long-ranged effects of 
the hydrodynamic interaction.
Our description corresponds to a
quasi two-dimensional suspension confined between parallel walls 
with spacing $h$.
The walls cut off the hydrodynamic forces on lateral scales much 
larger than $h$, making 
them effectively short range.
The {\em local} effects of the hydrodynamic interactions are, however,
accounted for by the concentration dependent mobility.
To describe the three dimensional problem we have to explicitly include the
the hydrodynamic velocity-field of the suspension in the description.
Although this could have far-reaching effects at small wavenumber, it should
be relatively unimportant for the statics of crystallization, whose physics
is dominated by wavenumbers near the structure-factor peak.

\section{phase transitions in a nonequilibrium problem}
\label{freezingphasetrans}

Since we are interested in the freezing transition,
we set up the equation of motion (\ref{freom}) for the density field, 
with a free energy functional that leads to a freezing 
transition in equilibrium.
The simplest of these is the Landau-Alexander-McTague free energy 
functional which we have discussed in section \ref{introfreezing}.
A spatially inhomogeneous density fluctuation $\phi (\br)$ 
(Fourier transform $\phik$) 
has, relative to the uniform fluid, a free energy
\cite{alexander,chailub}

\ber
\ft[\phik] &=& 
\frac{1}{2} \ik \frac{1}{\rho_0 \sok} \phik \phimk
\nonumber\\
&&
- \frac{1}{3!} b \ik \ip  \phimk \phip \phikmp
\nonumber \\
&&
+ \frac{1}{4!} c \ik \ip \ipp \times
\nonumber \\
&&
\quad\quad\quad
\phimk \phip {\phi}({{\bf p}'}) 
{\phi}({\bk - \bp -{\bf p}'}) 
\label{freeenalexmctague}
\eer
where $\sok$ is the equilibrium structure factor of the suspension.
The density correlation function at equilibrium calculated using
the above free energy is, to quadratic order,
\be
\lav \phikone \phiktwo \rav
=
\rho_0 \sok \delta(\bkone + \bktwo).
\ee
The equation of motion (\ref{freom}) becomes in reciprocal space
\subequations
\ber
&&\ddt \phik =
- \tpd \dt \rho_0 k^2  \frac{\delta \ft}{\delta \phimk} 
\nonumber\\
&&+ i \frac{\lambda}{\rho_0} k_z \ip \phip \phikmp + {\eta}(\bk),
\label{eomkwithfeom}
\\
&&
\lav \eta(\bk,t) \eta(\bkp,t') \rav 
=
\tpd 2 \dt k^2 \rho_0 \delta(\bk+\bkp) \delta(t-t').
\label{noisekspace}
\eer
\label{eomkwithf}
\endsubequations
Using the form (\ref{freeenalexmctague}) for the 
free energy, the equation of motion (\ref{eomkwithfeom}) becomes
\ber
\ddt \phik &=&
- \frac {\dt k^2}{\sok} \phik
+ i \frac{\lambda}{\rho_0} k_z \ip \phip \phikmp 
\nonumber
\\
&&+ {\eta}(\bk).
\label{eomnoneq}
\eer
where the contribution of anharmonic terms in the free energy
have not been put in.
This is because we are iterested in the {\em corrections} to the 
equilibrium bare correlation funtions due to the driving term.

It is easy to see that the additional nonlinear term
in equation (\ref{freom}) or (\ref{eomnoneq}) 
cannot be expressed in a form so as to be absorbed into an additional
term in the free energy.
This is done by checking that the functional curl of the term is 
nonvanishing.
We will call this  term in (\ref{eomnoneq}) the sedimentation
nonlinearity.

To describe phase transitions in this nonequilibrium problem,
we follow earlier work on critical phenomena in the presence of shear \cite{onuki} 
and shear-induced melting of colloidal crystals \cite{rrbagchi}. 
We assume that the Fokker-Planck equation that derives from the above 
Langevin equation (\ref{eomnoneq}) has a steady solution in
the long time limit, which we write as
\be
P[\phi] \sim e^{-{\Gamma [\phi]}}
\label{probss}
\ee
and expand the functional $\Gamma$ in a Taylor expansion
\be
\Gamma [\phi] = \half {\Gamma}_2 (1,2)\phi(1) \phi(2) 
+ \frac{1}{3!} {\Gamma}_3 (1,2,3) \phi(1) \phi(2) \phi(3)
+ \ldots
\ee
where $1,2$ etc. are arguments, for example spatial positions $x_1, x_2 $ etc.
The summation convention is implied for repeated arguments.
The $\Gamma_n$ are the vertex functions.
The calculation of moments from this probability functional gives,
for the connected parts, the relations \cite{fieldth}
\subequations
\begin{eqnarray}
\left< \phi(1) \phi(2) \right> & =
& {{\Gamma}_2 (1,2)}^{-1},
\label{corlsandvertex1}
\\
\left< \phi(1) \phi(2) \phi(3) \right> & =
- &  {\Gamma}_3(1',2',3') 
\left< \phi(1) \phi(1') \right> 
\times
\nonumber\\
&&
\left< \phi(2) \phi(2') \right>
\left< \phi(3) \phi(3') \right>,
\label{corlsandvertex2}
\end{eqnarray}
\label{corlsandvertex}
\endsubequations
and so on, which are the usual relations between the connected 
correlation functions and vertex parts.
The only difference here is that the steady state probability distribution
is not the inverse exponential of the equilibrium free energy functional.
The vertex functions $\Gamma_i$ above have to be therefore calculated
from the steady state probability distribution $P$.
However, the correlation functions
can be calculated from the Langevin equation itself by demanding
that the time derivative of all moments is zero, 
and these may be used in (\ref{corlsandvertex}) to calculate
the vertex functions upto any desired order, which gives the steady state
$P$ to the same order in the Taylor expansion.
For instance,
\subequations
\begin{eqnarray}
\left< \phikone \right. && \left. \phiktwo \right>  =
 {{\Gamma}_2 (\bkone,\bktwo)}^{-1},
\label{corlsandvertexk1}
\\
\left< \phikone \right. && \left. \phiktwo \phikthree \right> = 
\nonumber \\
&&
- \ikone \iktwo \ikthree 
{\Gamma}_3(\bkone',\bktwo',\bkthree') \times
\nonumber
\\
&&
\left< \phi(\bkone') \phikone \right> 
\left< \phi(\bktwo') \phiktwo \right>
\left< \phi(\bkthree') \phikthree \right>.
\label{corlsandvertexk2}
\end{eqnarray}
\label{corlsandvertexk}
\endsubequations
%Note that the $n^{\mbox{th}}$ vertex $\Gamma_n$ has dimensions of
%$V^n$ coming from the momentum integrals.
The second of these relations can be rewritten in terms of the
structure factor as
\be
\left< \phikone \phiktwo \phikthree \right>
 = - 
{\Gamma}_3(\bkone,\bktwo,\bkthree)
{\rho_0}^3 S(k_1)S(k_2)S(k_3).
\label{threeptvertexwithsk}
\ee

At thermal equilibrium, i.e., in the absence of driving, 
the vertex functions calculated to tree level
are simply the inverse correlation function 
and the higher vertices in the free energy:
\subequations
\begin{eqnarray}
{\Gamma}_2 (\bkone,\bktwo) 
&= & 
\frac {1}{\rho_0 S_0(k_1)} \delta(\bkone + \bktwo),
\label{corlsandvertexeq1}
\\
{\Gamma}_3 (\bkone,\bktwo,\bkthree) 
& = & -  b \, \delta (\bkone + \bktwo + \bkthree),
\label{corlsandvertexeq2}
\\
{\Gamma}_4 (\bkone,\bktwo,\bkthree,\bkfour) & = &
c \,  \delta (\bkone + \bktwo + \bkthree + \bkfour).
\label{corlsandvertexeq3}
\end{eqnarray}
\label{corlsandvertexeq}
\endsubequations
We calculate the perturbative correction to the 
vertices due to the sedimentation nonlinearity, from which we
obtain the steady state probability distribution  (\ref{probss}).
This $P$ can be treated as a mean-field probability distribution
which may be maximized to find the stable phase at any given temperature.
In other words, $\Gamma[\phi]$ can be treated as a nonequilibrium free energy.
Thus the perturbative corrections to the vertices can be interpreted
as additional terms generated by sedimentation in an effective free energy.

\section{perturbation theory}
\label{freezingperturbation}

We start with the linearized equation of motion 
(\ref{eomnoneq}) for the density field with
only the sedimentation nonlinearity,
\ber
\ddt \phik =
&-& \frac {\dt k^2}{\sok} \phik
\nonumber\\
&+& i \frac{\lambda}{\rho_0} k_z \ip \phip \phikmp + {\eta}(\bk),
%\nonumber
\eer
and calculate perturbative correction to the vertex functions.
The noise statistics is as given in (\ref{noisekspace})
which, for momentum labels $\bk$ and $\bmk$ reads
\be
\lav \eta(\bk,t) \eta(- \bk,t') \rav 
=
2 \dt k^2 \rho_0 V \delta(t-t')
\label{noisekspacewithvol}
\ee
where $V$ is the volume of the system. 
To the lowest (i.e. first) order, there is a correction to the 
three-point vertex, whereas the structure factor gets modified at
second order in P\'eclet number.
The four-point vertex ${\Gamma}_4$ will also have a correction  at 
second order, but we ignore this correction as this is simply a 
stabilizing term and small changes in it will not affect the 
physics of freezing significantly.
% (CHECK OKAY?).

In this section, we set up a hierarchy of equations for the moments
of $\phi$, which we close by decoupling the four-point function 
into a product of two two-point functions ( Hartree approximation).
Multiplying the equation of
motion (\ref{eomnoneq}) by $ \phik $, symmetrizing in $k,-k$
and setting time derivative of the two-point function to zero,
we obtain for {\em equal time} a relation between the two-point and the 
three-point functions:
\ber
2 \dddk && \lav \phimk  \phik  \rav =
\nonumber\\
i \frac{\lambda}{\rho_0} k_z &&
\ip [ \lav \phip \phikmp \phimk \rav - ( \bk \rightarrow - \bk) ]
\nonumber\\
&+& 2 N \dt k^2, 
\label{twopointequaltime}
\eer
where we have used the equal-time correlation between the noise and the
density fluctuation,
\be
\lav \phi (-\bk,t) {\eta} (\bk,t) \rav
= N \dt k^2, 
\label{noiseandensityequaltime}
\ee
which is proved in detail in the appendix.
Here $N = \rho_0 V$ is the total number of particles in the system.
Since the correlation function is symmetric under interchange of 
$\bk$ and $\bmk$, 
(\ref{twopointequaltime}) can be rewritten as
\ber
\dddk && \lav \phimk  \phik  \rav =
\nonumber\\
&&
\lll \ip  \lav \phip \phikmp \phimk \rav 
+ N \dt k^2, 
\label{twopointequaltimeb}
\eer
Similarly, multiplying with two fields $\phi$ at different wavevectors
but equal times, we obtain a relation involving the third and 
fourth moments:
\ber
\dt && \lav \phikone \phiktwo \phikthree \rav \si \kisqbysk 
=
\nonumber\\
&i& \frac{\lambda}{\rho_0}  k_{1z} \ip \lav \phip \phi 
( \bkone -\bp) \phiktwo \phikthree \rav
\nonumber\\
&& \quad   {\mbox{et cycl.}}
\label{hier2}
\eer
%where the word cyclic stands for two other terms 
%formed by cyclic permutations of the first.
To obtain this equation we have  used the fact that the correlation
of two fields $\phi$ and one noise $\eta$ vanishes at equal time
(see appendix).
If we proceed in this manner we will keep obtaining higher-order correlations
in terms of the lower ones.
We are, however, interested only in the lowest-order correlations, so
we close the hierarchy by replacing the four-point function by a 
product of two two-point functions. Only pairings of $p$ with 
$k_2$ or $k_3$ give nonzero contributions, because of the $k_{1z}$ 
prefactor, so the dummy index $\bp$ on the right hand side of
(\ref{hier2}) can only take the values $- \bktwo$ and $- \bkthree$.
The two terms thus obtained are identical, and we arrive at the following
relation between the three-point and two-point functions:
\ber
\dt \si && \kisqbysk
\lav \phikone \phiktwo \phikthree \rav
=
\nonumber\\
&&2 i {\lambda} \rho_0 \delta ( \bkone + \bktwo + \bkthree )
S_0(k_1) S_0(k_2) S_0(k_3)
\si \kizbysk.
\nonumber\\
\label{threeptfnhartree}
\eer
Using this in combination with (\ref{threeptvertexwithsk}) 
we obtain the correction to the three-point vertex from sedimentation:
\be
\delta {\Gamma}_3 (\bkone, \bktwo, \bkthree)
=
- \frac{2 i \lambda}{\dt {\rho_0}^2} \,
\frac{\displaystyle \si \kizbysk}{\displaystyle \si \kisqbysk}
\delta ( \bkone + \bktwo + \bkthree ).
\label{correctiontogamma3}
\ee
Combining (\ref{twopointequaltime}) with (\ref{threeptfnhartree}),
we obtain the correction to the structure factor as an integral over 
wavevectors.
In the limit of infinite volume, the sum over wavenumbers  
goes from $0$ to an upper cutoff determined by the smallest length
scale in the problem. Our coarse-grained model breaks down on scales 
of order the particle radius $R$; 
the upper cutoff is therefore $\frac{2 \pi}{R}$.
Non-dimensionalizing the momentum argument in the integral with the 
$\frac{1}{\rho_0}$ outside, i.e. by multiplying all momenta by 
the average interparticle spacing $\sep$, 
we finally obtain
\ber
S(\bk) &=& \sok 
- 2 \/  {\left(  \frac{\lambda \sep}{\dt} \right) }^2 \sok
\frac{k_z}{k^2}
\nonumber\\
&& 
\times 
{\int}_0^{\frac{2 \pi \sep}{R}} \frac{d^d p}{{(2 \pi)}^d} \,
\sok \, \sop \, \sokmp \quad
\nonumber \\
&&
\times
\left\{
\frac
{\displaystyle
\frac{p_z}{\sop} 
- \frac{k_z}{\sok} 
- \frac{p_z - k_z}{\sopmk} 
}
{\displaystyle
\frac{p^2}{\sop} 
+ \frac{k^2}{\sok} 
+ \frac{{(\bp - \bk)}^2}{\sopmk} 
}
\right\},
\label{skintegral}
\eer
where {\em all } the wavevectors in the above equation, including the 
$k$'s multiplying the integral, are non-dimensionalized by multiplying
with $\sep$.

Note that the upper cutoff and hence the integral goes to $\infty$ in the
limit $ R \longrightarrow 0$.
Let us calculate how the structure factor correction behaves in this
limit.
The integrand, for large arguments, $\sim \frac{1}{p}$ 
(note that $S(p) \longrightarrow 1$ for large $p$), 
so by power counting, the integral $\sim R^{(1-d)}$.
This apparent divergence is however controlled by the powers of
$R$ in the P\'eclet number, which was $ R^{2 d}$.
We are thus left with a correction that $\sim R^{(1+d)}$, which vanishes
as $R \longrightarrow 0$ for all $d$.
This is consistent with the fact that in this limit there are no
hydrodynamic interactions.

The results of this section can also be obtained using diagrammatic
perturbation theory \cite{lahiriphdfr}.
The correlation function calculated to one loop without self consistency
and the three point vertex at tree level give results identical to
those obtained above.

\section{the structure factor}
\label{freezingstrucfac}

\begin{figure}
\epsfxsize=8cm 
\centerline{\epsfbox{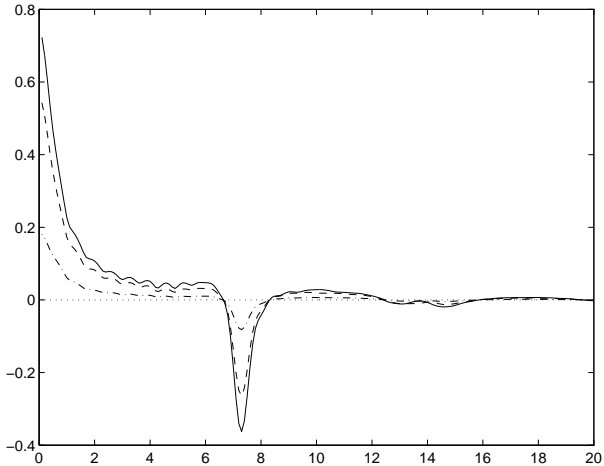}}
\caption{
The correction $\frac{S(k) - \sok}{\sok}$  as a function of $k$
for different angles 
with respect to the sedimentation direction, 
in two dimensions:
$\theta = 90^o$ (dotted line) showing no change;
$\theta = 60^o$ (dash-dot);
$\theta = 30^o$ (dashed line);
and $\theta = 0^o$ (continuous line).
Clearly, the correlations are reduced near the peak, and the
reduction is the largest along the sedimentation direction.
}
\label{fig2ddeltaskvk}
\end{figure}

We computed the structure factor of the sedimenting suspension by numerically
computing the integral in Eq. (\ref{skintegral})
\cite{fnfreezingintegral}.
We used a 40-point Gaussian quadrature routine 
\cite{numrec} and took the
large $k$ cutoff of the integral, $\frac{2 \pi a}{R}$ to be 20,
corresponding to $\frac{\sep}{R} \sim 3$.
The integral has reflection symmetry about $\theta = \pi / 2$, so that
the structure factor at an angle $\alpha$ from the horizontal
($\theta = \pi / 2 - \alpha$) 
is the same as that at an angle  $- \alpha $ from the horizontal
($\theta = \pi / 2 + \alpha$).
This follows from the invariance of the integral
under $k_z \longrightarrow -k_z$.

\begin{figure}
\epsfxsize=8cm 
\centerline{\epsfbox{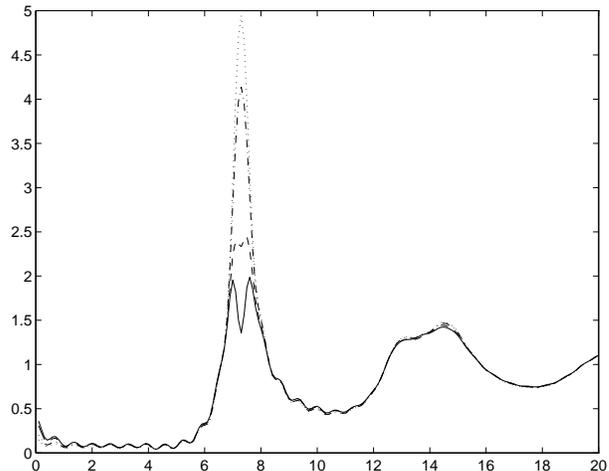}}
\caption{
$S(k)$ for a sedimenting colloidal suspension in two dimensions
as a function of $k$, at 
$Pe = 1 $ for four different angles
with respect to the sedimentation direction:
$\theta = 90^o$ (dotted line)
which is the same as the undistorted structure factor;
$\theta = 60^o$ (dash-dot);
$\theta = 30^o$ (dashed line);
and $\theta = 0^o$ (continuous line).
}
\label{figskvk2d}
\end{figure}

The numerical results are presented in Figs. 
\ref{fig2ddeltaskvk} and \ref{figskvk2d}.
The first shows the fractional change in the structure factor per unit P\'{e}clet
number as a function of wavenumber and angle.
The {\em reduction}  happens to be a maximum at the peak itself,
and is a monotonic function of the angle 
between the horizontal (where it is maximum) and the vertical (where it
goes to zero).
In Fig. \ref{figskvk2d}  we show the numerically computed structure factor
in two dimensions.
The equilibrium structure factor has been taken from Monte Carlo 
simulations of a two-dimensional system of particles
interacting via a screened Coulomb potential, 
just above freezing \cite{cdas}.
The peak of the equilibrium $S(k)$ was at $k=7.3$ (which is close
to the average interparticle separation $k_0 \sim 2 \pi$
in the units we are using) and the peak value was $S_0(k_0)=4.947$.
We have chosen the P\'eclet number $ \frac{\lambda a}{\dt} = 1\/$
for these plots.
This is a rather large number, and our calculation, which goes only
upto leading order in perturbation theory is probably not accurate
for such large driving. 
The large value has been chosen to make the deviations from the 
equilibrium structure easily visible.

It is important to note that the split-peak 
in $S(k)$ is a high P\'eclet number effect.
Since the correction to $S(k)$ is peaked at the same place as $S(k)$ itself,
there can be {no shift} in the position of the peak of  the extremum.
At low P\'eclet number, therefore, 
the structure factor has a global maximum at $k \sim 7.3$, which becomes a 
local minimum at a sufficiently large $Pe$. 
At what value of $Pe$ this happens depends on the angle,
the required driving going to infinity as the horizontal direction 
is approached.

\begin{figure}
\epsfxsize=8cm 
\centerline{\epsfbox{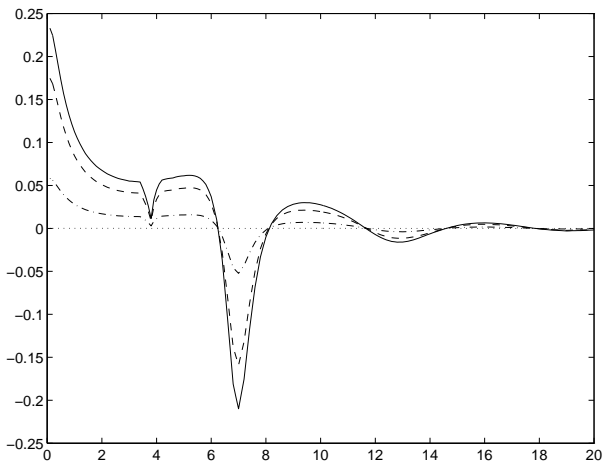}}
\caption{
The correction $\frac{S(k) - \sok}{\sok}$  as a function of $k$
for different angles with respect to the sedimentation direction, 
in three dimensions;
$\theta = 90^o$ (dotted line) showing no change;
$\theta = 60^o$ (dash-dot);
$\theta = 30^o$ (dashed line);
and $\theta = 0^o$ (continuous line).
As in the two dimensional case, 
the correlations are reduced near the peak, and the
reduction is the largest along the sedimentation direction.
}
\label{fig3ddeltaskvk}
\end{figure}

\begin{figure}
\epsfxsize=8cm 
\centerline{\epsfbox{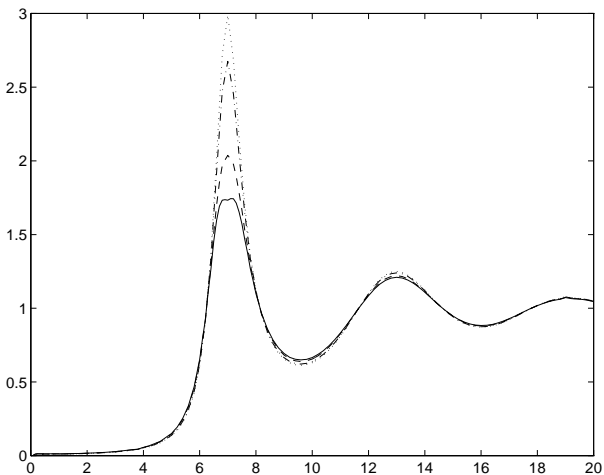}}
\caption{
The three dimensional 
$S(k)$ as a function of $k$ after sedimentation correction, at 
$Pe = 1$ for different angles with respect to the sedimentation direction:
$\theta = 90^o$ (dotted line) $\equiv$ the undistorted structure factor;
$\theta = 60^o$ (dash-dot);
$\theta = 30^o$ (dashed line);
and $\theta = 0^o$ (continuous line).
}
\label{fig3dskvk}
\end{figure}

We have studied the freezing transition only in two dimensions
(section \ref{freezingfreezing}), but we present structure factor
data for three dimensions as well
( Figs. \ref{fig3ddeltaskvk} and \ref{fig3dskvk}). 
In three dimensions the integral in (\ref{skintegral}), and also 
the structure factor correction, have an azimuthal symmetry.
In addition there is the 
reflection symmetry about $\theta = \pi / 2$ as in the two-dimensional case.
For this case we used a 30-point Gaussian quadrature routine \cite{numrec},
and checked
that the answer does not change significantly between 28-point and 30-point.
The structure factor after the sedimentation correction is shown 
in Fig. \ref{fig3dskvk}.
For the equilibrium structure factor we have used Monte Carlo results
for a $\frac{1}{r^6}$ liquid \cite{hansenschiff}, made continuous by  suitable 
interpolation.
The P\'eclet number is once again $Pe = 1$.

\section{freezing}
\label{freezingfreezing}
The effect of the corrected three-point vertex as well as the structure
factor correction have to be considered while describing the liquid-solid
transition in the presence of sedimentation. 
The latter has a relatively simple effect: since $S(k)$ is reduced in 
all directions, there  is a tendency towards melting.
The correction (\ref{correctiontogamma3}) to the three-point vertex,
being imaginary, has more interesting implications.
Combining (\ref{correctiontogamma3}) with (\ref{corlsandvertexeq2})
we conclude that the nonequilibrium effects shift the cubic vertex,
\be
B \longrightarrow B 
-\frac{2 i}{\rho_0} \left( \frac{ \lambda \sep}{\dt} \right) \,
\frac{\displaystyle \si \kizbysk}{\displaystyle \si \kisqbysk},
\ee
where we have scaled all wavevectors by multiplying with the average
interparticle separation $\sep$.
With reference to (\ref{landauparameters}), we can find the cubic term in the 
Landau expansion for the nonequilibrium system,
\be
- b \longrightarrow - b 
- (2 p n^{-1/2})
\frac{2 i \lambda}{\dt } \,
\frac{\displaystyle \si \kizbysk}{\displaystyle \si \kisqbysk}.
\label{bchange}
\ee

Introducing a real dimensionless quantity which is first order 
in P\'eclet number,
\be
\beta_1 =
2 \left( \frac{\lambda a}{\dt } \right)  \,
\frac{\displaystyle \si \kizbysk}{\displaystyle \si \kisqbysk},
\label{beta1}
\ee
we rewrite (\ref{bchange}) as 
\be
b \longrightarrow b +  i (2 p n^{-1/2}) \beta_1
\label{bchange2}
\ee
The correction to the Landau parameter $b$ is entirely imaginary,
and makes the effective $b$ complex.
The free energy is a real, positive quantity, so a complex coefficient 
$B$ in \ref{flandauintro2} would imply that the complex Fourier
amplitudes $\phik$ pick up phases.
In equilibrium, when $B$ is real (and positive by construction),
the phases of $\phimk, \phip$ and $\phikmp$ add up 
to $2 \pi$ so as to make the cubic term as negative as possible.
In other words, if we write the amplitude as 
\be
\phik = |\phik| e^{i \theta(\bk)}
\ee
then 
\ber
{\cal{\theta}} ( \bkone,\bktwo,\bkthree )
& = & \theta(\bkone)+ \theta(\bktwo)+ \theta(\bkthree) 
\nonumber
\\
& = & 2 \pi
\eer
for sets of vectors that form triangles, i.e. satisfy $\bkone + \bktwo + \bkthree = 0$.
Note that the cubic term is always real, whatever the 
phases of the wavevectors, since for any set
\{$\bkone,\bktwo,\bkthree$\} 
there is another with all wavevectors reversed which has exactly the opposite
total phase ${\cal{\theta}}$.
However the contribution to the free energy from two such terms is
$2 cos {\cal{\theta}}$, which should be $ = 2 $ to minimize the 
free energy.

\begin{figure}[htbp]
\begin{center}
\leavevmode
\psfig{figure=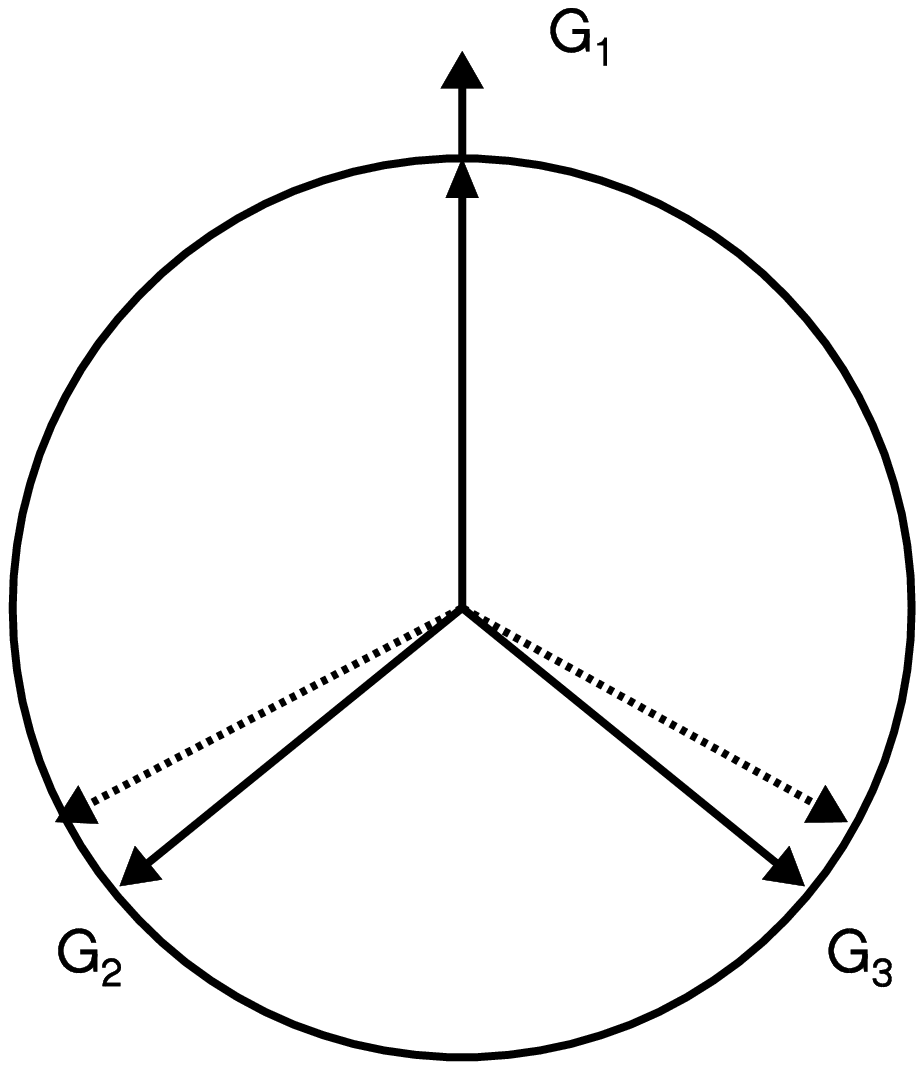,width=6cm}
\psfig{figure=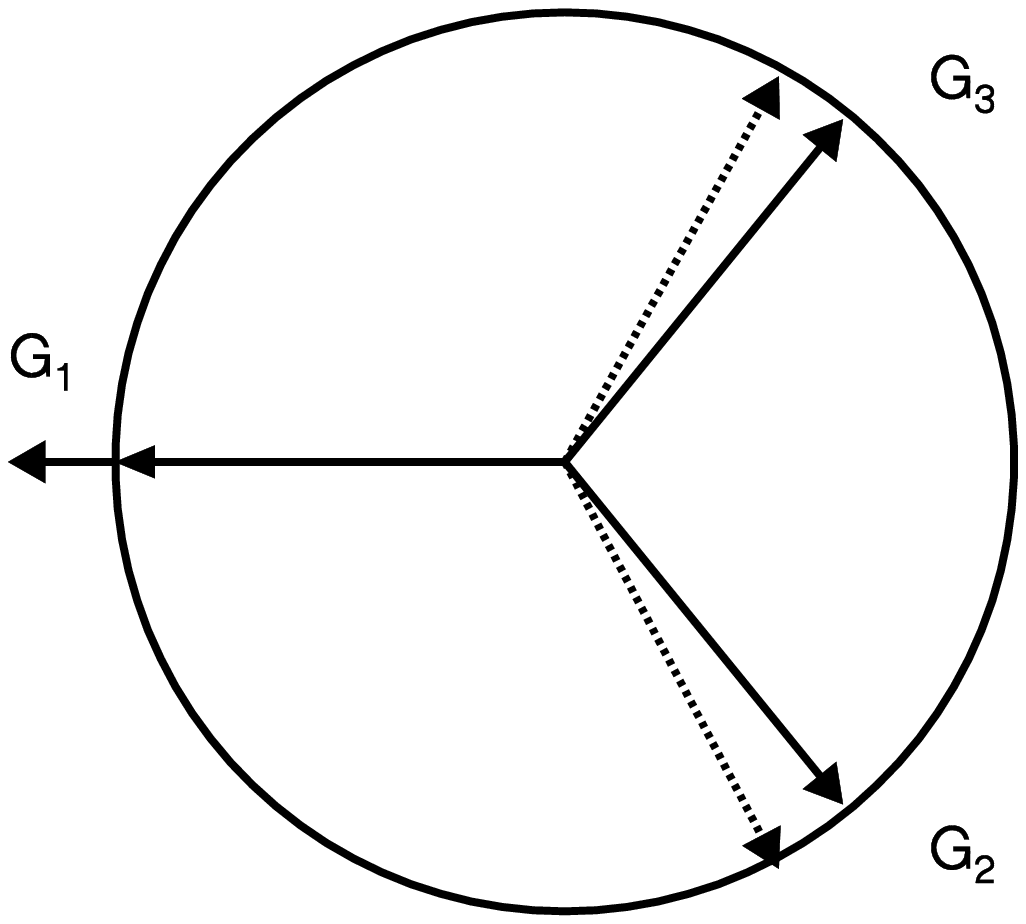,width=6cm}
\caption{
Possible orientations of inner shell 
reciprocal lattice vectors in a sedimenting
colloidal crystal.
The continuous lines stand for the distorted crystal, and the 
dotted lines for the original undistorted lattice.
The sedimentation direction is downwards on the plane of the
paper, i.e., antiparallel to $\bgone$ in the first figure.
}
\label{figdistperp}
\end{center}
\end{figure}

With a complex coefficient as in (\ref{bchange}), the free energy
is minimized if the phases satisfy 
\be
{\cal{\theta}} ( \bkone,\bktwo,\bkthree )
+ {\mbox {arctan}} \left( \frac{(2 p n^{-1/2}) \beta_1}{b} \right)
= 2 \pi.
\ee
This `phase shift' in the order  parameters may be visualized in
real space as a shift of one set of lattice planes relative to its
equilibrium position.
Note that scattering experiments measure only the squared modulus
${|\phik|}^2$, and will not see a phase shift. 

Freezing is determined by the {\em modulus} of the modified
coefficient of the cubic term, which is 
\be
b \longrightarrow \sqrt{b^2 +  4 p^2 n^{-1} {\beta_1}^2}.
\ee
The prefactor multiplying $\/{\beta_1}^2$ is a number that depends on
the type of lattice.
For a triangular lattice in two dimensions, 
$n = 6$ (six reciprocal lattice vectors in the first shell)
and
$p=1$ (each reciprocal lattice vector belongs to one of the two triangles).
Thus
\be
b \longrightarrow \sqrt{b^2 +  \frac{2}{3} {\beta_1}^2} 
\quad \quad
\mbox{(for a planar triangular lattice)}.
\label{modbcorrection}
\ee
Clearly, the magnitude of $b$ can only increase due to sedimentation.
Thus, as far as this term is concerned, sedimentation seems 
to favour ordering.
This effect has to compete with the reduction of the structure factor
and only a quantitative analysis can decide their  combined effect.

In the rest of this section we present results for the simplest
case, that of a planar triangular lattice.
We have described the equilibrium freezing with a one-order parameter
Landau theory, choosing the Landau coefficients to match the 
structure factor peak height at coexistence for our data, i.e.
$S_0(k_0) = 4.947$.
Our Landau parameters were $b= 6, c=59.364$ and, at coexistence,
%sriram
$a = a_0 = 0.20214271$, giving a polynomial
\be
F = 0.10107 a r^2 - r^3 + 2.5 r^4 
\ee
where $r$ is the reduced order parameter, defined in  (\ref{opr}).  

Note that the correction $\beta_1$ [eqn. (\ref{beta1})] 
vanishes if the reciprocal
lattice vectors remain on the shell $|{\bf G}| = G_0$, and the
structure factor remains unchanged.
This follows from (\ref{bchange}) and the fact that the sum of $k_i$
has to be zero.
Thus, to benefit from the free energy reduction that an increase
in $b$ can cause, the lattice may like to distort, thus sending
one of the vectors out of the shell while keeping the sum equal 
to zero by rotating the other two towards or away from each
other.
We have considered two such distortions in our analysis, as shown
in Fig. \ref{figdistperp},
which  seem to be most likely from the symmetry of the problem.
It is also imaginable that the lattice distorts to an asymmetric
state with two degenerate minima of the free energy;
we have not made a complete scan of all possible distortions
to rule out such possibilities, but looked at only the above
two on grounds of plausibility.

Such distortions apart, the structure factor is itself affected by 
the driving, so $\beta_1$ is nonzero even without 
pushing the reciprocal lattice
vectors out of the shell.
The undistorted crystal is therefore  also a strong candidate.
The crucial issue is the relative importance of the two effects:
$\beta_1$ and $S(k)$.
Though the correction to the three-point vertex is first order in
$Pe$, the correction to $b$ for small $Pe$ goes as ${Pe}^2$  
as can be seen simply by binomially expanding the square-root 
in (\ref{modbcorrection}).
For this reason, combined with the fact that the numerical 
value of $\beta_1$ is rather small at modest driving rates, 
the $\Gamma_3$ effect is negligible compared to the
structure factor reduction.
For the Landau potential chosen, we have found numerically that the
structure factor reduction makes the free energy function 
go unstable well before $b$ can change significantly.

\begin{figure}
\epsfxsize=8cm 
\centerline{\epsfbox{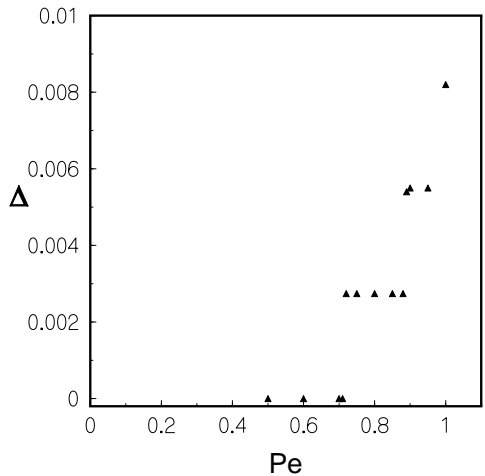}}
\caption{
The distortion $\Delta$ as a function of P\'eclet number.
}
\label{figdeltavpe}
\end{figure}

The physics, therefore, is dominated entirely by the structure factor.
This implies that sedimentation will neccessarily lead to 
a shift of the liquid-solid phase boundary towards the solid
side.
In other words, since the correlations are reduced,
a system which was at liquid-solid coexistence in the absence of sedimentation
will choose the {\em liquid} phase if sedimentation is switched on.

The reduction of the structure factor peak height 
is maximum along the sedimentation direction and
falls monotonically as the angle with respect to the horizontal
is decreased, going to zero in the horizontal direction.
It is therefore expected that one of the wavevectors will point
along the $x$ direction.
This immediately rules out the possibility shown in Fig. \ref{figdistperp}(a)
in favour of that shown in Fig. \ref{figdistperp}(b).
Even if there is no distortion, this orientation is the best among the
undistorted ones.
We now ask: is there any reason the lattice may wish to distort
due to structure factor effects alone?

Consider the distortion sketched in Fig. \ref{figdistperp}(b) --
the reciprocal lattice vector $G_1$ growing slightly larger,
\be
|{\bf G_1}|  = G_0 \/ (1 + \Delta),
\ee
and the others rotating towards the $x$ axis so as to keep the
sum of the vectors zero.
If the P\'eclet number is not too large, $S(k)$ continues to have a peak at
$k_0$ even in the presence of sedimentation
(section \ref{freezingstrucfac}), so that
there is a free energy cost to pay for the expanding $G_1$ which
is moving off the peak. 
The other two, however, have moved closer to the
horizontal, i.e. in a direction such that the driving affects
them less, and the peak height at these wavevectors is higher
than it would have been had they not rotated at all.
There is a competetion between these two effects, and its not
{\em a priori} clear which will win, 
so we computed the free energy of distorted crystals at coexistence
to find the most favourable value of the distortion or angle.
Note that in any case the liquid is the favoured phase, but
we can still search among all the crystalline phases and find which 
of them has the lowest free energy.
This, as we demonstrate shortly, gives
us some hint as to what to expect at other temperatures.

We found that for low P\'eclet numbers the undistorted crystal
is favoured, but at around $Pe = 0.72$ a distorted crystal 
becomes favourable (Fig. \ref{figdeltavpe}).
There seems to be a first order jump in what we might call the
distortion order parameters $\Delta$.
This is not surprising, as we may regard the nonequilibrium free energy
coming out of our analysis as a function of $\Delta$, whose
power series expansion will have odd order terms since $\Delta \longrightarrow
-\Delta$ is not a symmetry of the problem.

The distorted crystal appears at rather high P\'eclet number
(Fig. \ref{figdeltavpe}), certainly well beyond the range of 
validity of our perturbation theory.
The reason for the sudden distortion is probably related to the 
original peak $k_0$ becoming a minimum and two peaks developing
on either side (Fig. \ref{figskvk2d}).
The several plateaus that appear at even higher $Pe$ are even more
intriguing.
The intriguing results regarding split peaks
(Figs. \ref{figskvk2d},\ref{fig3dskvk})
and distortions are however somewhat speculative in the 
absence of a reliable theory for large driving.

We have also compared the configurations shown in Figs 
\ref{figdistperp}(a),
and \ref{figdistperp}(b) 
and found, as expected, that the latter has a lower free energy at all 
values of the driving.
Thus one immediate consequence of our theory is that the crystal,
when it forms, will form with the orientation shown 
in Fig. \ref{figdistperp}(b),
i.e. with lattice planes parallel to the sedimentation direction.
This agrees with intuition, since there are shocks 
\cite{kynch,brewer,vansarlooshuse}
travelling along the sedimentation direction.

\begin{figure}
\epsfxsize=8cm 
\centerline{\epsfbox{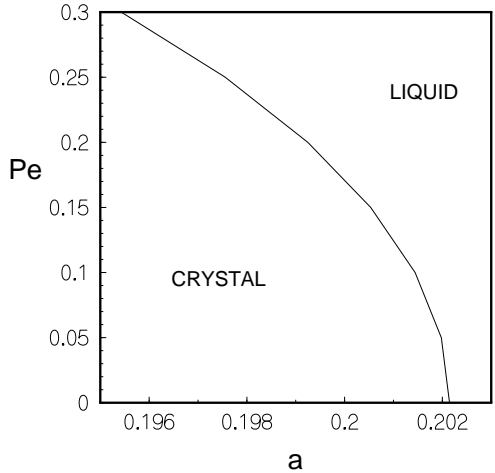}}
\caption{
Nonequilibrium phase diagram of the sedimenting colloidal suspension.
The `$a$' on the abcissa is the inverse structure 
factor height in the liquid phase of the suspension.
}
\label{sedpdg}
\end{figure}

Finally, one would like to look at the system below 
the equilibrium freezing temperature $T_c$,
and find how the melting P\'eclet number depends on $T-T_c$.
Strictly speaking this requires structure factor data in the supercooled
liquid, which one should use in (\ref{skintegral}) to compute
the corrected $S(k)$.
Such data, however, is not readily available, and in particular
for the case of the planar triangular lattice that we are considering,
it is very hard to sustain a supercooled liquid for too long in a 
simulation with monodisperse systems.

We therefore attempt the following approximate theory.
Near $T_c$, the structure factor is that at coexistence plus 
a perturbative correction.
The critical P\'eclet  number $Pe_c$ at which the crystal melts
is an object which is itself perturbative in $T-T_c$.
In calculating $Pe_c$, if we neglect the perturbative corrections
to $S_0(k)$, we will be making an error only at a higer order.
We can therefore legitimately do a Landau theory at a  $T$ below but
close to $T_c$ using the structure factor {\em at coexistence}.
The $S(k)$ corrections will all be calculated using this $S_0(k;T_c)$,
the temperature entering only in the Landau parameter 
%sriram 
$a$.
The result is shown in Fig. \ref{sedpdg}.

\section{concluding remarks}
\label{sec:conclude}

We have shown that a uniform drift, as in steady-state sedimentation, 
reduces translational correlations in a suspension of interacting Brownian
particles, as a result of local hydrodynamic interactions 
that reduce the mobility of a dense region.
This reduction is largest in the direction of sedimentation.
At sufficiently high P\'eclet number, the first peak
of the structure factor splits due to the fact that the
reduction is strongest where the initial correlation was largest. 

For the freezing transition we find that
the phase boundary in a charge stabilized colloidal
suspension shifts towards the crystal side.
The shape of the phase boundary is quadratic 
in $Pe$ as shown in Fig. \ref{sedpdg}.

Let us now make some estimates for P\'eclet numbers in realistic
experimental situations to see whether the values of $Pe$ in 
Fig. \ref{sedpdg} are accessible.
The P\'eclet number for a dense suspension, at the scale of the particle 
radius, is given by Eq.  \ref{peclet3}, which can be rewritten as
\be
Pe  = 
\left(\frac{D_0}{\dt} \right)
\left(\frac{\sep}{R} \right)
{\left[ \frac{{\partial^2} \left( {\frac{v}{v_0}} \right)} 
{\partial {\lr}^2} \right]}_{\r0}\/
\left(\frac{mgR}{k_B T} \right),
\label{pecletinremarks}
\ee
provided the hindered settling speed (or the mobility) is roughly 
linear in the volume fraction in the range of interest.
Taking the specific case of polystyrene spheres of diameter $15.5 \mu m$
in water  
\cite{xue}, we note that this is true for volume fraction
between $0.1$ and $0.3$ \cite{xuefig}.
The volume fraction used in our calculations
($\frac{\sep}{R} \sim 3$, from the cutoff used for the integrals,
i.e. volume fraction $\frac{\rho}{\rho_0} \sim 0.16$)
falls within this range.
>From the data presented in \cite{xuefig}, the term within the
square brackets in (\ref{pecletinremarks}) is found to be of order $1.5$.
%\ref{mutwodef}

The first term $\left(\frac{D_0}{\dt} \right)$ in (\ref{pecletinremarks})
is the ratio of the bare to the collective diffusion constant, 
which has a value of about $10$ near the freezing transition of
a colloidal suspension \cite{lowen}.
The ratio of the interparticle spacing to the radius of the polyballs,
$\left(\frac{\sep}{R} \right)$, is $\sim 3$.
The product of the first three factors in \ref{pecletinremarks} 
is thus $10 \times 3 \times 1.5 = 45$.
Finally, we have to estimate the bare P\'eclet  number $\frac{mgR}{k_B T}$.
This turns out to be $\sim 0.5$ for polystyrene \cite{fnbarepeclet}.
The P\'eclet number for the interacting suspension is therefore $22.5$. 
This is rather large, which means that our lowest-order theory 
is probably not reliable here. It does however suggest 
that the effect of steady-state sedimentation on freezing 
should be large for this system. 

In general, in choosing a system on which to test our results, 
two competing needs must be kept in mind. On the one hand, the 
ordering length scale should be large compared to the particle 
size, so that the simple concentration-dependence of the mobility  
(see the remarks towards the end of section \ref{freezingeomsection})  
in our model is valid. On the other hand, the particle size itself  
should be large enough to give an appreciable Peclet number. 

A comparison with shear induced melting is tempting, since a perturbation
theory similar in spirit to the present approach \cite{rrbagchi}
is known to give a shift in the phase boundary that is 
quadratic in the shear rate.
There are essential differences however:
firstly the {\em only} significant effect of shear is to suppress
correlations, whereas for a sedimenting suspension, there are
nontrivial corrections to the three point vertex.
The effect of these latter corrections happens to be negligible on
the freezing transition as we have shown, making the phase boundary
shift in a manner similar to the shear case.
Secondly, the sheared colloidal crystal at low shear-rates is 
a slightly distorted crystal essentially stuck in a statically 
sheared configuration.
At low shear therefore (where perturbation theory is valid), the
crystal is essentially in an  equilibrium state, and its properties
can be derived from an equlibrium picture \cite{rlshear}.
For uniform driving, however, there are nonzero currents even
at low P\'eclet numbers, making it a more genuinely nonequilibrium 
problem.

An interesting feature of our theory is that the sign of 
$\frac{\partial \mu}{\partial \rho}$ does not figure in the results.
Thus even if we have a suspension whose settling speed {\em increases}
due to many body effects, our theory will still predict a reduction
in the correlations and a consequent shift in the phase boundary
towards the crystal side.

Finally, a few caveats.  In \cite{sedlat} we have shown that a sedimenting
colloidal crystalline lattice is likely to be unstable at long 
wavelength. The theory proposed in this paper is probably 
nonetheless applicable to the formation of the crystalline phase 
at least at low P\'eclet numbers, since the physics 
of freezing is centred primarily around the ordering wavenumber. 
The coupling of concentration and shear (tilt) 
considered in \cite{sedlat,crow}
may also be important in the liquid state. 
The detailed consequences will be different because shear
stresses have a finite lifetime in the liquid, in contrast
to that in a solid where it lasts for ever if the strain is present.
The effect of the sedimentation instability is thus very different
in a liquid.
This is an additional aspect that we have totally ignored in  the above
treatment, and one that a more sophisticated approach must address.
However, further development of the theory should be motivated
by experiments, which we eagerly await.

\appendix
\section{field-noise correlations at equal time}
Consider a set of Langevin equations of the form
\subequations
\ber
{\dot \phi_{\alpha}}  
&=& 
f({ \phi})
+\eta_{\alpha},
\label{fraplang}
\\
\left< 
\eta_{\alpha} (t)
\eta_{\beta} (t')
 \right> 
&=&  
2  k_B T \/ {\Gamma}_{\alpha \beta} 
\delta(t-t').
\label{frapnoise}
\eer
\label{fraplangandnoise}
\endsubequations
where ${ \phi}$ stands for $\{\phi_1, \phi_2, \ldots \phi_n \}$,
and $f$ can be any arbitrary function, in general nonlinear.
We shall calculate the field-noise correlations using
Novikov's theorem
\cite{novikov,mamazenko,frzinn},
which states that for any functional of the noise $A[\eta]$, 
\be
\lav A \eta_\alpha(t) \rav
=
2 T \Gamma_{\alpha \beta} \lav \frac
{\delta A}
{\delta \eta_\beta(t)}.
\rav
\label{novikovthm}
\ee
We first calculate the average of the product of one field with one
noise, which by (\ref{novikovthm}) is
\be
\lav \phi_\alpha(t) \eta_\lambda(t') \rav 
=
\lav 
\frac
{ \delta \phi_\alpha(t)}
{\delta \eta_\lambda(t') } 
\rav. 
\ee
The equation (\ref{fraplang}) can be formally solved to give
\be
\phi_\alpha(t) = 
\int_{t_0}^t dt'  f\left( { \phi}(t') \right)
+
\int_{t_0}^t dt' \eta_\lambda(t')
\ee
where the $\eta_\lambda$ dependence in the 
first term is present through $\phi(t)$,
which depends on the noise amplitudes at all past times but not those of
the future.
The derivative
\be
{\frac
{ \delta \phi_\alpha(t)}
{\delta \eta_\lambda(t') }
}
\ee
vanishes for $t<t'$ by causality. To calculate the equal time value
of this derivative, let us calculate it first for $t=t'+\epsilon$
for infinitesimal positive $\epsilon$.
For $t>t'$,
\be
\frac
{\delta
\phi_\alpha(t)} 
{\delta
\eta_\lambda (t')}
=
\int_{t_0}^t dt''  \frac{\partial f}{\partial \phi_\beta} (t'')
\frac{\delta \phi_\beta(t'')}{\delta \eta_\lambda(t')} \quad + 1.
\ee
For $t=t'$, all the arguments $t''$ in the integral are
times before $t'$, so the functional derivative inside the integral,
and hence the integral itself vanishes.
Thus the forward derivative is simply equal to $1$.
The backward derivative is zero, therefore
\be
\frac
{\delta
\phi_\alpha(t)} 
{\delta
\eta_\lambda t'}
=
\theta(t-t').
\ee
At equal time this is discontinuous, but we recall that the delta
function noises are nothing but limits of Gaussians, so the theta
function is not really a discontinuous function at the origin, but rather 
rises smoothly from $0$ to $1$ taking the midpoint value $\half$ at the 
origin. Thus
\be
\frac {\delta \phi_\alpha(t)} {\delta \eta_\lambda (t)} = \half,
\ee
which, combined with (\ref{novikovthm}) gives
\be
\lav \phi_\alpha(t) \eta_\lambda(t') \rav 
=
k_B T \Gamma_{\alpha \lambda}.
\ee
This can be shown with more rigour by starting from a Gaussian distribution,
calculating the equal time correlation and then taking the width of the
Gaussian  to zero \cite{frzinn}.
In section \ref{freezingperturbation}
we have used an analogous result for reciprocal space variables.

Let us now consider the product of two fields and a noise.
Again, using (\ref{novikovthm}), we have
\be
\lav \phi_\alpha(t) \phi_\beta(t) \eta_\lambda(t) \rav 
=
\lav 
\phi_\beta(t) \frac { \delta \phi_\alpha(t)} {\delta \eta_\lambda(t) } 
+
\phi_\alpha(t) \frac { \delta \phi_\beta(t)} {\delta \eta_\lambda(t) } 
\rav 
\ee
where we have used the chain rule for functional derivatives.
Now each of the partial derivatives $= \half$, therefore
\be
\lav \phi_\alpha(t) \phi_\beta(t) \eta_\lambda(t) \rav 
=
\half \/
\lav \phi_\beta(t) + \phi_\alpha(t) \rav. 
\ee
In section \ref{freezingperturbation} we do a perturbation theory about the
liquid phase, calculating liquid state correlations, so the first moments
of $\phi$ are all zero, thus making all averages of the form
$ \langle \phi \phi \eta \rangle$ vanish.

For completeness we
explicitly  check the vanishing of the first moment in the presence
of driving. 
Averaging (\ref{eomnoneq}) over noise, and setting the 
time derivative of the field average
to zero (steady state), we have
\be
 \frac {\dt k^2}{\sok} \lav \phik \rav
=
 i \frac{\lambda}{\rho_0} k_z \ip \lav \phip \phikmp \rav. 
\label{firstmomentaverage}
\ee
The second moment on the right is nonzero only if $-p = k-p$, i.e.
$k=0$, so that the first moment  $\lav \phik \rav$ vanishes for all
$k \neq 0$.
But since $\phi$ is a deviation variable, its $k=0$ component is
zero by construction.

\acknowledgements
We would like to thank Mustansir Barma,
Pinaki Majumdar, Rajaram Nityananda and Rahul Pandit for discussions and 
Sushan Konar for the use of interpolation programs.
A special word of thanks is due to Chinmay Das who supplied structure
factor data from his Monte Carlo simulations \cite{cdas}.

Most of the work was done when R.L. was at the Physics Department,
Indian Institute of Science, Bangalore.
The facilities of the institue are acknowledged, including
the the computational resources of the Supercomputer Education
and Research Centre. 
R.L. was supported by CSIR (India) during the same period.


\begin{references}
\bibitem{fnfreezingframes}
Correlation  functions calculated at unequal times do have a signature
of the reference frame. 
This should in principle be observable in dynamic light scattering 
experiments.
\bibitem{fnfreezingintegral}
Strictly speaking, the integral should be performed in a region 
such that both $\bk$ and $\bkmp$ are within a sphere of radius
$\Lambda =\frac{2 \pi a}{R}$.
We have, for simplicity, simply integrated over all $\bk$
inside the sphere, i.e. our calculation covers a somewhat larger
area of integration than the correct one.
However, in the limit of large, though
finite $\Lambda$, the error incurred by this is negligible.
\end{references}
\end{document}